\documentclass[preprint,eqsecnum,aps,amssymb,tightenlines]{revtex4}
\usepackage{latexsym}

\newcommand{\be}{\begin{equation}}
\newcommand{\ee}{\end{equation}}
\newcommand{\bea}{\begin{eqnarray}}
\newcommand{\eea}{\end{eqnarray}}
\newcommand{\ba}{\begin{array}}
\newcommand{\ea}{\end{array}}
\newcommand{\nn}{\nonumber \\}

\begin{document}

\title{Noncommutative Gravity \vspace{.3cm}}

\author{E. Harikumar and Victor O. Rivelles}

\affiliation{Instituto de F\'\i sica, Universidade de S\~ao Paulo\\
 Caixa Postal 66318, 05315-970, S\~ao Paulo, SP, Brazil\\
E-mail: hari@fma.if.usp.br and rivelles@fma.if.usp.br\\ \vspace{2cm}}

\vspace{2cm}

\begin{abstract}
We consider simple extensions of noncommutativity from flat to
curved spacetime. One possibility is to have a generalization of the
Moyal product with a  covariantly constant 
noncommutative tensor $\theta^{\mu\nu}$. In this case the spacetime
symmetry is restricted to volume preserving diffeomorphisms which also
preserve $\theta^{\mu\nu}$. Another possibility is an extension of the
Kontsevich 
product to curved spacetime. In both cases the noncommutative product
is nonassociative. We find the order $\theta^2$ noncommutative
correction to the Newtonian  potential in the case of a
covariantly constant $\theta^{\mu\nu}$. It is still 
of the form $1/r$ plus an angle dependent piece. The coupling to
matter gives rise to a propagator which is $\theta$ dependent. 
\end{abstract}

\maketitle

%\newpage

\section{Introduction}

Noncommuting coordinates were proposed a long time ago as a
generalization of the phase space of quantum mechanics
\cite{Snyder:1946qz}. The 
idea was resurrected in the string theory context when a decoupling
limit was found that yields a noncommutative gauge theory induced by
the Moyal product \cite{Seiberg:1999vs}. From then on its properties were
studied in detail and a large amount of information on noncommutative
field theories were gained \cite{reviews}. 

The main characteristic of noncommutative field theories is that it
incorporates  
non-local effects in a controllable way. This is reminiscent of its
stringy origin where the gravitational sector was decoupled but still
left some traces through the noncommutativity. In the quantum theory
particles present a kind of dipole structure \cite{Seiberg:2000gc} while in
quantum field theory there appears a mixing of ultraviolet and
infrared divergences \cite{Minwalla:1999px}. Another interesting effect is that
matter fields can feel noncommutativity as a background gravitational
field \cite{Rivelles:2002ez}.  

So there is a good understanding of noncommutative effects in matter
and gauge fields in flat spacetime. The next step is to incorporate
gravity and thus consider curved spacetimes. The main trouble we have
to face now is how to  
deal with the noncommutative parameter which is usually taken to be
constant in flat spacetime. This gave rise to a large amount of work
to study deformations of general relativity
\cite{Szabo:2006wx,unclassified,chamseddine,wess}. Usually noncommutative 
deformations of gravity lead to a complex metric and tangent space
groups larger than the Lorentz group \cite{chamseddine}. Another
approach is to replace the diffeomorphism invariance of general
relativity by a twisted version of it \cite{wess}.
Both of these approaches lead to gravitational theories
which are not simple extensions of general relativity. An attempt in
this direction considers that the enveloping
algebra of the Lorentz group is gauged and volume preserving
diffeomorphisms are used instead of general coordinate
transformations. Gravitational theories based on volume preserving
diffeomorphisms are usually known as unimodular gravity theories 
\cite{Alvarez:2005iy} and a noncommutative deformation of it is
proposed in \cite{Calmet:2005qm}. 

More recently an attempt to derive noncommutative gravity from string
theory was 
presented \cite{Alvarez-Gaume:2006bn}. The dynamics of closed strings
in the presence of a constant B-field induces a gravitational action
in the next-to-leading order in the Seiberg-Witten limit
\cite{Seiberg:1999vs} and some of the three gravitons interaction
vertices are derived. It happens that these vertices can not be
obtained from an action written only with Moyal products and the authors
claim that this is due to the fact that the Moyal product changes
under a space-time diffeomorphism. 

In this paper we present an alternative approach. Instead of
implementing diffeomorphisms in the noncommutative setting we extend the
properties of noncommutativity in 
flat spacetime to curved spacetime in a way that is as simple as
possible. In flat spacetime we usually assume 
that the noncommutative parameter $\theta^{\mu\nu}$ is constant and
its infinitesimal functional variation vanishes. As we shall see, we
need only to assume that the functional variation of $\theta^{\mu\nu}$ is zero
since its constancy will be a consequence of preserving translation
invariance. We then assume the same condition in curved spacetime. Now
we obtain that $\theta^{\mu\nu}$ is a covariantly constant tensor
\cite{Lizzi:2002ib} and that the symmetries of 
spacetime reduce to volume preserving diffeomorphisms which also 
preserve $\theta^{\mu\nu}$. An alternative extension to curved spacetime is to
consider the flat spacetime Kontsevich product where $\theta^{\mu\nu}$ is no
longer constant but satisfies the Jacobi identities. This
condition can easily be extended to curved spacetime.  

In our setting noncommutativity can couple to the spacetime
geometric tensors. It also manifests itself through the coupling to matter
by extensions of the Moyal and Kontsevich products to curved
spacetime. While in flat spacetime both products are associative, no
simple extension of them to curved spacetime is found which also
preserves associativity. 

We study the consequences of such a proposal finding noncommutative
corrections to the Newtonian potential in the case where $\theta^{\mu\nu}$ is
covariantly constant. The correction still has the Newtonian form
$1/r$ but with an effective Newton constant. In addition, there is a
piece which has an angular dependence.  

Section II discusses how we can extend the properties of
the noncommutativity parameter $\theta^{\mu\nu}$ from flat spacetime
to curved spacetime. We consider the case of constant
$\theta^{\mu\nu}$ used in the Moyal product and the case of a non
constant $\theta^{\mu\nu}$ used in the Kontsevich product. In Section
III we discuss the coupling to matter and how the nonassociativity
emerges. Section IV is devoted to the study of the coupling of
noncommutativity to gravity in the case where  $\theta^{\mu\nu}$  is
covariantly constant. In the next Section we discuss the linearized
limit to find the noncommutative correction to the Newtonian
potential. Finally, the last Section is devoted to some comments and
conclusions. 

\section{Noncommutativity in flat and curved spacetime}

In flat spacetime noncommutativity is described by a constant
antisymmetric matrix $\theta^{\mu\nu}$ which has the same value in all
inertial frames and leads to a breakdown of Lorentz
invariance. Let us consider this in more
detail in order to be able to find which properties from flat
spacetime can be generalized to a curved one. Let us promote 
$\theta^{\mu\nu}$ to a tensor under Lorentz transformations and perform
a general coordinate transformation with infinitesimal parameter $\xi^\mu$
\be
\label{2.1}
\delta \theta^{\mu\nu} = \xi^\lambda \partial_\lambda \theta^{\mu\nu} -
\partial_\lambda \xi^{\mu} \theta^{\lambda\nu} + \partial_\lambda
\xi^{\nu} \theta^{\lambda\mu}.
\ee
Recall that $\delta \theta^{\mu\nu}$ stands for a functional variation
of $\theta^{\mu\nu}$ at the same point and we are going to require its
vanishing. For a rigid translation $\xi^\mu$ is
constant and we get $\partial_\lambda \theta^{\mu\nu} = 0$ so that all
components of $\theta^{\mu\nu}$ are constant. For a Lorentz
transformation $\xi^\mu = 
{\Lambda^\mu}_\nu x^ \nu$ we get 
\be
\label{2.2}
{\Lambda^\mu}_\lambda \theta^{\lambda\nu} - {\Lambda^\nu}_\lambda
\theta^{\lambda\mu} = 0.
\ee
For instance, if the only nonvanishing component of $\theta^{\mu\nu}$
is $\theta^{12}$ then 
Lorentz boosts in $3$-direction and rotations in the $1-2$ plane are
still preserved as well as translations in any direction. 

There is one more transformation which solves $\delta
\theta^{\mu\nu}=0$ and that is
\be
\label{2.6}
\xi^\mu = \theta^{\mu\nu} \partial_\nu \xi,
\ee
where $\xi$ is a scalar function. Notice that $\partial_\mu \xi^\mu
= 0$. If we work with curvilinear coordinates the volume element $d^4
x$ is a scalar density of weight 1 so that it is invariant under an
infinitesimal transformation satisfying $\partial_\mu \xi^\mu=
0$. Then this is a volume preserving transformation. However,
transformation (\ref{2.6}) is a specific volume preserving
transformation. Since the commutator of two infinitesimal
transformations with parameters $\xi^\mu_1$ and $\xi^\mu_2$ gives a
transformation with parameter $\xi^\mu_3 = \xi^\lambda_1 \partial_\lambda
\xi^\mu_2 - \xi^\lambda_2 \partial_\lambda \xi^\mu_1$, for the
particular case (\ref{2.6}) we get 
\be
\label{2.6a}
\xi^\mu_3 = \theta^{\mu\nu} \partial_\nu \xi_3, \qquad \xi_3 = -
\theta^{\mu\nu} \partial_\mu \xi_1  \partial_\nu \xi_2, 
\ee
so that they form a subgroup of the volume preserving
transformations. Since $\theta^{\mu\nu}$ is preserved by these
transformations we get a sympletic subgroup of the volume preserving
diffeomorphisms \cite{Jackiw:2002pn}. These 
residual symmetries are not useful in flat spacetime, even though
there are some attempts to use them \cite{Chaichian:2004qk}. However, they
will be useful in curved spacetime as it will be shown. Hence, the
requirement that $\theta^{\mu\nu}$ be a tensor whose functional
variation under 
Poincar\'e transformations vanishes,  seems to be a good starting
point for generalization. Notice that $\delta \theta^{\mu\nu}=0$ leads
not only to a condition on $\theta^{\mu\nu}$ but it also implies that Lorentz
transformations in some directions are broken. 
 
Consider now a spacetime whose Riemannian geometry is associated to
a metric tensor $g_{\mu\nu}$. Under an infinitesimal diffeomorphism an
antisymmetric tensor $\theta^{\mu\nu}$ transforms as
\be
\label{2.3}
\delta \theta^{\mu\nu} = \xi^\lambda D_\lambda \theta^{\mu\nu} -
D_\lambda \xi^\mu \theta^{\lambda\nu} + D_\lambda \xi^\nu
\theta^{\lambda\mu}.
\ee
We now impose that $\delta \theta^{\mu\nu}=0$, as in flat spacetime,
and we will solve this equation as constraints on $\theta^{\mu\nu}$
and on the diffeomorphisms. A solution which generalizes the constancy
of $\theta^{\mu\nu}$ in flat spacetime is that it is covariantly
constant 
\be
\label{2.4}
D_\lambda \theta^{\mu\nu} = 0, 
\ee
which implies that 
\be
\label{2.5}
D_\lambda \xi^\mu \theta^{\lambda\nu} - D_\lambda \xi^\nu
\theta^{\lambda\mu} = 0.
\ee
The solution to this last equation is simply (\ref{2.6}). Notice that
$D_\mu \xi^\mu = 0$ so that there is a residual symmetry similar to
what happens in flat spacetime. We are left with the sympletic subgroup of
the volume preserving diffeomorphisms which also preserves a
covariantly constant 
$\theta^{\mu\nu}$. We would like to point out that this symmetry also
appears in another formulation of noncommutative gravity which makes
use of noncommuting coordinates \cite{Calmet:2005qm}.

Another starting point would be the use of the Kontsevich product
\cite{Kontsevich:1997vb} instead of the Moyal product. In this case
$\theta^{\mu\nu}$ is a tensor even in flat spacetime and satisfies 
the Jacobi identities
\be 
\label{2.7}
\theta^{\lambda\mu} \partial_\lambda \theta^{\nu\rho} + 
\theta^{\lambda\nu} \partial_\lambda \theta^{\rho\mu} + 
\theta^{\lambda\rho} \partial_\lambda \theta^{\mu\nu} = 0,
\ee
which turns the Kontsevich product into an associative product. What
is remarkable is that besides respecting Poincar\'e symmetry
$\theta^{\mu\nu}$ is left invariant by the sympletic subgroup of
volume preserving diffeomorphisms for $\theta^{\mu\nu}$ satisfying
(\ref{2.7}). So the Kontsevich product also allows a volume preserving
diffeomorphism if $\theta^{\mu\nu}$ is form invariant. Even though 
providing an alternative way to implement noncommutativity in flat 
spacetime its consequences in quantum field theories are far less understood
than in the Moyal case \cite{Das:2003kw}. 

Using the Kontsevich product we no longer have the condition $\delta
\theta^{\mu\nu}=0$ as in flat spacetime. The generalization to curved
spacetime is straightforward since (\ref{2.7}) is covariantized as   
\be 
\label{2.8}
\theta^{\lambda\mu} D_\lambda \theta^{\nu\rho} + 
\theta^{\lambda\nu} D_\lambda \theta^{\rho\mu} + 
\theta^{\lambda\rho} D_\lambda \theta^{\mu\nu} = 0.
\ee
This condition now replaces (\ref{2.4}) and we find no further
condition on general diffeomorphisms. As in the flat spacetime case we
can require that $\theta^{\mu\nu}$ is form invariant and 
we find that volume preserving diffeomorphisms are allowed if use is
made of (\ref{2.8}).

\section{Coupling to Matter}

We assume that the geometric quantities are unaffected by
noncommutativity so that 
we can define the Christoffel symbol, covariant derivatives, the
Riemann tensor, and so on in 
the usual way. Noncommutativity can appear as in flat spacetime, by
replacing the ordinary product by a Moyal product where the ordinary
derivatives have been replaced by covariant ones. For two tensors
$T_{\mu_1\dots}$ and $S_{\nu_1\dots}$ we write formally 
\be
\label{3.1}
T_{\mu_1\dots}(x) \star S_{\nu_1\dots}(x)  = e^{ \frac{i}{2}
  \theta^{\alpha\beta} D_\alpha(x) D_\beta(y) } T_{\mu_1\dots}(x)
S_{\nu_1\dots}(y) |_{y=x}.
\ee
This is a formal definition because the covariant derivatives do not
commute and we must specify a precise ordering for second and higher
order terms in an expansion in $\theta$. 

One of the main properties of the Moyal product in flat spacetime is
associativity so we would like to retain it in curved
spacetime. Unfortunately we could not find any ordering which makes
the Moyal product in (\ref{3.1}) associative. For instance, defining
(\ref{3.1}) up to second order in $\theta$ as 
\be
\label{3.2}
T \star S = T S + \frac{i}{2} \theta^{\alpha\beta} D_\alpha T \, D_\beta S -
\frac{1}{8} \theta^{{\alpha_1}{\beta_1}} \theta^{{\alpha_2}{\beta_2}}
D_{{\alpha_1}} D_{\alpha_2} T \,\, D_{\beta_1} D_{\beta_2} S,
\ee
(tensorial indices were omitted) we find that 
\bea
\label{3.3}
(T \star S) \star U &-& T \star (S \star U) = 
- \frac{1}{4} \theta^{{\alpha_1}{\beta_1}}
\theta^{{\alpha_2}{\beta_2}} \left( - \frac{1}{2} D_{{\alpha_1}} T
  \,\, D_{\alpha_2} S \,\, [D_{\beta_1},D_{\beta_2}]U   \right.\nn
&+& \left. \frac{1}{2}
[D_{\beta_1},D_{\beta_2}] T \,\, D_{\alpha_2} S \,\, D_{{\alpha_1}} U + 
D_{{\alpha_1}} T \,\,  [D_{\beta_1},D_{\beta_2}] S \,\, D_{\alpha_2} U 
\right).
\eea
Even if we change the ordering in the quadratic term in $\theta$ in
(\ref{3.2}) we were not able to recover associativity. For instance,
if we take the symmetric part in ${{\alpha_1}}{{\alpha_2}}$ and also in
${\beta_1}{\beta_2}$ we still get the same result (\ref{3.3}). Other
modifications of (\ref{3.2}), like having terms with one derivative in
S and three derivatives in T, were also tried without success. 

Another possibility would be to consider the Kontsevich product which
in flat spacetime is associative if (\ref{2.7}) holds. In curved
spacetime we generalize the Kontsevich product by replacing ordinary
derivatives by covariant ones and up to second order in $\theta$ we
obtain   
\bea
\label{3.4}
T \star S &=& T S +  \frac{i}{2} \theta^{\alpha\beta} D_\alpha T \,
D_\beta S - \frac{1}{8} \theta^{{\alpha_1}{\beta_1}} 
\theta^{{\alpha_2}{\beta_2}} D_{{\alpha_1}} D_{\alpha_2} T \,\,
D_{\beta_1} D_{\beta_2} S  - \nn
&-&  \frac{1}{12} \theta^{\alpha_1\beta_1}
D_{\beta_1} \theta^{\alpha_2\beta_2} \left( D_{\alpha_1} D_{\alpha_2}
  T \, D_{\beta_2} S + D_{\beta_2} T \, D_{\alpha_1} D_{\alpha_2} S
\right).
\eea
An explicit calculation shows that if we use (\ref{2.8}) one still
gets (\ref{3.3}) so the proposed extension of the Kontsevich
product to curved spacetime is no 
longer associative. 

The origin of nonassociativity can be traced to string theory. When we
consider D-branes in curved backgrounds it is known that a 
nonassociative star product is needed \cite{Cornalba:2001sm}. There is
nothing wrong with nonassociative 
products. It just turns life more difficult. We have to be extremely
careful when writing products since the order now must be clearly
specified.  

In flat spacetime the Moyal product of two fields gives the ordinary
product of them plus total derivative terms. When writing an action, 
terms in $\theta$ can then be disregarded. In curved spacetime this 
property no longer holds for the Moyal product (\ref{3.2}). It gives a
total (covariant) derivative term plus $\theta$ contributions
involving curvature terms. So when 
writing an action which is quadratic in the fields the Moyal product
gives non trivial contributions involving $\theta$. For instance, we can
write the action for a scalar field as
\be
{S}=\frac{1}{2}\int \sqrt{-g} d^4 x \left
  [g^{\mu\nu}\partial_\mu\Phi \star \partial_\nu\Phi - m^2 \Phi \star 
  \Phi\right], 
\ee
and expanding as in (\ref{3.2}) to second order in $\theta$ we find 
\bea
{S} = \frac{1}{2}\int \sqrt{-g} d^4 x \left[ g^{\mu\nu}
  \partial_\mu \Phi \partial_\nu \Phi
-\frac{1}{8} g^{\mu\nu} \theta^{\alpha_1\beta_1} \theta^{\alpha_2\beta_2} 
D_{\alpha_1} D_{\alpha_2} \partial_\mu \Phi D_{\beta_1} D_{\beta_2}
\partial_\nu\Phi\right. \nonumber\\
\left.- m^2 \Phi \Phi + \frac{m^2}{8} \theta^{\alpha_1\beta_1}
  \theta^{\alpha_2\beta_2} D_{\alpha_1} \partial_{\alpha_2} \Phi
  D_{\beta_1} \partial_{\beta_2}\Phi
\right]. 
\eea
Notice that in flat spacetime the propagators receive no $\theta$
correction because the quadratic terms in the action have no $\theta$
dependence. The noncommutative contribution appears exclusively at
the interaction vertices. In curved spacetime we find that the
propagators are now $\theta$ dependent and are modified by 
noncommutativity. This is in striking difference with flat spacetime. 

%where we have terms to second order in $\theta$. The $\Phi$ field equation of
%motion following from the above is
% \bea
% -g^{\mu\nu}D_\mu D_\nu\Phi -m^2 \Phi\Phi
% +\frac{1}{16}\theta^{\alpha_1\beta_1}\theta^{\alpha_2\beta_2}
% g^{\mu\nu}D_\mu D_{\alpha_2}\left( R_{{\alpha_1}{\beta_1}{\beta_2}}^{~~~~~~\rho} D_\rho D_\nu\Phi +
%  R_{{\alpha_1}{\beta_1}{\nu}}^{~~~~~~\rho} D_{\beta_2} D_\rho\Phi
% \right)\nonumber\\
% -\frac{1}{16} \theta^{\alpha_1\beta_1}\theta^{\alpha_2\beta_2} 
% D_{\alpha_2}\left(R_{{\alpha_1}{\beta_1}{\beta_2}}^{~~~~~~\rho} D_\rho\Phi\right)=0.
% \eea
% Here we note that the NC parameter $\theta$ get coupled to Riemann
% tensor also. 

\section{Noncommutative Gravity}

Since we are leaving the geometry untouched by noncommutativity the
only way to study its effect in a purely gravitational context is
through the coupling to the geometric tensors. Recall that $\theta^{\mu\nu}$ is
a tensor but has no dynamics. It is constrained either to be
covariantly constant (\ref{2.4}) or by (\ref{2.8}). For simplicity we
will consider the case where $\theta^{\mu\nu}$ is covariantly constant and
consider only the lowest order couplings in $\theta^{\mu\nu}$.  Then it   
couples to the scalar curvature as $ \theta^{\mu\nu}
\theta_{\mu\nu} R$, to the Ricci tensor as $\theta^{\mu\nu}
{\theta_{\nu}}^\lambda R_{\mu\lambda}$, and to the Riemann tensor as
$\theta^{\mu\nu} {\theta^{\alpha}}_\beta {R_{\mu\nu\alpha}}^\beta$ (the
coupling $\theta^{\mu\nu} {\theta^{\alpha}}_\beta
{R_{\mu\alpha\nu}}^\beta$ reduces to the former one). Since
noncommutativity is very small we can assume a perturbative scheme for
$\theta^{\mu\nu}$ and look for corrections to general relativity. If we
consider only vacuum solutions of Einstein equations then the first
two couplings are not relevant and only the third one should be
regarded. 

Then we start with the Einstein-Hilbert action plus the noncommutative
contribution 
\be
\label{4.1}
S_{NC} = \frac{1}{16\pi} \int d^4 x \,\, \sqrt{-g} \,\, \theta^{\mu\nu}
{\theta^{\alpha}}_\beta {R_{\mu\nu\alpha}}^\beta.
\ee
Notice that we need a dimensionfull coupling constant with dimension
$L^{-6}$ which can be absorbed in $\theta$, so that it has dimension
$L^{-1}$ instead of the usual dimension $L^2$. There is also a global sign
which can not be absorbed in $\theta$ but which can be changed by
$\theta^2 \rightarrow - \theta^2$. 

As discussed in section II, the introduction of a covariantly constant
$\theta^{\mu\nu}$ reduces the diffeomorphism transformations to a subgroup of
volume preserving transformations. 
As remarked in the introduction, gravity theories based on volume
preserving transformations are known as unimodular gravity
theories \cite{Alvarez:2005iy}. Since $D_\mu \xi^\mu = 0$ for a volume
preserving 
transformation then the determinant of the metric, which transforms 
as a scalar density of weight $-2$, is invariant so that $\det g$ is
constant and usually taken to be minus one. Then we have a theory with an
absolute spacetime volume element, the modulus. Unimodular gravity has
many interesting properties. The cosmological constant appears as an
integration constant and the action has a finite polynomial form in the
metric \cite{vanderBij:1981uw}. Since we can take any solution of general
relativity and go to a frame where $\det g=-1$ there is no physical
difference between both gravity theories
\cite{Alvarez:2005iy,Finkelstein:2000pg}.  In our case, the
Einstein-Hilbert action is invariant under general diffeomorphisms but
$S_{NC}$ (\ref{4.1}) is invariant under the restricted class of volume
preserving diffeomorphisms. 

The equations of motion we obtain are then
\be
\label{4.2}
\frac{1}{G} \left( R_{\mu\nu} - \frac{1}{2} g_{\mu\nu} R \right) -
\frac{1}{2} g_{\mu\nu} \theta^{\rho\lambda} {\theta^\alpha}_\beta
{R_{\rho\lambda\alpha}}^\beta - \frac{1}{2} \theta^{\alpha\beta}
\theta_{\mu\gamma} {R_{\alpha\beta\nu}}^\gamma - \frac{1}{2}
\theta^{\alpha\beta} \theta_{\nu\gamma} {R_{\alpha\beta\mu}}^\gamma =
0, 
\ee
and since the Einstein tensor is covariantly conserved this implies that 
\be
\label{4.3}
3 \theta^{\alpha\beta} \theta^{\rho\sigma} D_\alpha
R_{\mu\beta\rho\sigma} - 2 \theta^{\alpha\beta} {\theta^\rho}_\mu
D_\alpha R_{\beta\rho} = 0.
\ee
As we will see, this last equation, together with (\ref{2.4}), puts 
stringent constraints on the components of $\theta^{\mu\nu}$.

\section{Linearized Noncommutative Gravity}

To study the effects of noncommutativity on the gravitational field we
will consider the linearized approximation and obtain corrections to
the Newtonian potential. As usual we expand the metric around flat
spacetime, $g_{\mu\nu} = \eta_{\mu\nu} + h_{\mu\nu}$. The linearization
of (\ref{4.2}) gives
\bea
\label{5.1}
&& \frac{1}{G} \left[ \Box h_{\mu\nu} + \partial_\mu \partial_\nu
  {h_\rho}^\rho - \partial^\rho \partial_\mu h_{\nu\rho} -
  \partial^\rho \partial_\nu h_{\mu\rho} - \eta_{\mu\nu} \left( \Box
    {h_\rho}^\rho - \partial^\rho \partial^\sigma h_{\rho\sigma}
    \right) \right] \\
&+&\theta^{\alpha\beta} \theta_{\mu\gamma} \left(
            \partial_\nu \partial_\alpha {h_\beta}^\gamma -
            \partial^\gamma \partial_\alpha h_{\beta\nu} \right) +
          \theta^{\alpha\beta} \theta_{\nu\gamma} \left( 
            \partial_\mu \partial_\alpha {h_\beta}^\gamma -
            \partial^\gamma \partial_\alpha h_{\beta\mu} \right) + 
2 \eta_{\mu\nu} \theta^{\alpha\beta} {\theta^\rho}_\sigma
\partial_\rho \partial_\alpha {h_\beta}^\sigma = 0, \nonumber
\eea
while (\ref{2.4}) yields
\be 
\label{5.2}
\partial_\mu \theta^{\alpha\beta} + \frac{1}{2} \left( \partial_\mu
  {h_\lambda}^{[\alpha} +  \partial_\lambda 
  {h_\mu}^{[\alpha} - \partial^{[\alpha} h_{\mu\lambda} \right)
\theta^{\lambda\beta]} = 0.
\ee
In unimodular gravity we have $\det g_{\mu\nu} = -1$ which, at the
linearized level, implies that $h_{\mu\nu}$ is traceless. Then
  (\ref{5.1}) gives
\be
\label{5.3}
\Box h_{\mu\nu} - \partial^\rho \partial_\mu h_{\nu\rho} -
  \partial^\rho \partial_\nu h_{\mu\rho} + \eta_{\mu\nu} \partial^\rho
  \partial^\sigma h_{\rho\sigma} = {\cal O}(\theta^2),
\ee
and contracting this equation with $\eta^{\mu\nu}$ we get 
\be
\label{5.4}
\partial^\mu \partial^\nu h_{\mu\nu} = {\cal O}(\theta^2),
\ee
so that (\ref{5.3}) reduces to
\be
\label{5.5}
\Box h_{\mu\nu} - \partial^\rho \partial_\mu h_{\nu\rho} -
  \partial^\rho \partial_\nu h_{\mu\rho} = {\cal O}(\theta^2).
\ee

The first step is to find the solution for the Newtonian potential in
unimodular gravity. We can either take the solution in Cartesian
coordinates 
\be
\label{5.6}
g_{\mu\nu} = \eta_{\mu\nu} + \delta_{\mu\nu} h, \qquad h =
\frac{-2GM}{r},
\ee
where $r=\sqrt{x^2 + y^2 + z^2}$, and perform a coordinate
transformation so that $\det g_{\mu\nu} = -1$, or solve (\ref{5.4})
and (\ref{5.5}) directly with a vanishing rhs. The coordinate
transformation is given by  
\be
\label{5.7}
x^{\prime 0} = x^0, \qquad x^{\prime i} = \left( 1 - \frac{h}{2}
\right) x^i.
\ee
Anyway, we find that the Newtonian potential in Cartesian coordinates
is given by
\be
\label{5.8}
h_{00} = h, \qquad h_{0i} = 0, \qquad h_{ij} = n^i n^j h,
\ee
where $n^i = x^i/r$ is the radial unit vector. As required, $h_{\mu\nu}$ is
traceless and satisfies $ \partial^\mu \partial^\nu h_{\mu\nu} = 0$. 

Having (\ref{5.8}) as a solution of (\ref{5.1}) to order zero in $\theta$
and first order in $GM$ we
can now solve (\ref{5.2}) perturbatively in $\theta$ and $GM$. Then
$\theta^{\mu\nu}$ must satisfy $\partial_\mu \theta^{\alpha\beta}=0$, that is,
$\theta^{\mu\nu}$ is constant. 

We can now find the $\theta$ dependent terms which contribute to
$g_{\mu\nu}$. We write
$g_{\mu\nu} = \eta_{\mu\nu} + h_{\mu\nu} + \overline{h}_{\mu\nu}$ 
where $h_{\mu\nu}$ is given by (\ref{5.8}) and $\overline{h}_{\mu\nu}$
is of order $\theta^2$ and traceless. We then find that (\ref{5.3})
reduces to 
\bea
\label{5.9}
&& \frac{1}{G} \left( \Box \overline{h}_{\mu\nu} - \partial^\rho \partial_\mu
\overline{h}_{\nu\rho} - \partial^\rho \partial_\nu
\overline{h}_{\mu\rho} + \eta_{\mu\nu} \partial^\rho \partial^\sigma
\overline{h}_{\rho\sigma} \right) =  \\
&& - \theta^{\alpha\beta} \theta_{\mu\gamma} \left(
            \partial_\nu \partial_\alpha {h_\beta}^\gamma -
            \partial^\gamma \partial_\alpha h_{\beta\nu} \right) - 
          \theta^{\alpha\beta} \theta_{\nu\gamma} \left( 
            \partial_\mu \partial_\alpha {h_\beta}^\gamma -
            \partial^\gamma \partial_\alpha h_{\beta\mu} \right) - 
2 \eta_{\mu\nu} \theta^{\alpha\beta} {\theta^\rho}_\sigma
\partial_\rho \partial_\alpha {h_\beta}^\sigma. \nonumber
\eea

We have also to consider the equation coming from (\ref{4.3}). Since
it is second order in $\theta$ there is no term with $\overline{h}$
and we get 
\be
\label{5.10}
3 \theta^{\alpha\beta} \theta^{\rho\sigma}
\partial_\nu \partial_\rho \partial_\alpha h_{\beta\sigma} +
\theta^{\alpha\beta} \theta_{\nu\gamma} \left( \Box \partial_\alpha 
  {h_\beta}^\gamma - \partial^\gamma \partial^\mu \partial_\alpha
  h_{\beta\mu} \right) = 0. 
\ee
Using the field equation for $h_{\mu\nu}$ it reduces to
\be
\label{5.11}
\theta^{\alpha\beta} \theta^{\rho\sigma}
\partial_\nu \partial_\rho \partial_\alpha h_{\beta\sigma} = 0.
\ee
For $\nu=0$ the equation is trivial while for $\nu\not=0$ (\ref{5.11})
yields
\be
\label{5.12}
\theta^{0j} \theta^{0k} \partial_i \partial_j \partial_k h_{00} +
\theta^{jm} \theta^{kn} \partial_i \partial_j \partial_k h_{mn} = 0.
\ee
Introducing the vector notation $\theta^{0i} = \tilde{\theta}^i,\quad
\theta^{ij} = \epsilon^{ijk} \theta^k$, we find the
solution of (\ref{5.12}) to be 
\be
\label{5.12a}
\vec{\tilde{\theta}} = \pm \vec{\theta}. 
\ee
Then the time-space components of $\theta^{\mu\nu}$ are
proportional to the space-space components. This is to be contrasted
with the situation of noncommutative quantum field theories in flat
spacetime. There the theories with time-space components have troubles
with causality. This happens because there is no decoupling limit in
string theory which generates a background with a B-field which has
different electric and magnetic components which are perpendicular to
each other \cite{Seiberg:2000gc}. In our case, the string
theory background has a B-field with electric and magnetic components
which are parallel or antiparallel. In this case the same happens, no
decoupling limit does exist \cite{Aharony:2000gz}. However, we are now
in curved spacetime and closed strings need no longer be
decoupled. 

We can then go back to (\ref{5.9}) and solve it. We get 
\bea
\label{5.13}
\overline{h}_{00} &=& 3 G \vec{\theta}^2 h - 
G (\vec{n} \cdot \vec{\theta})^2 h + \alpha \vec{\theta}^2 h, \nn
\overline{h}_{0i} &=& 0, \nn
\overline{h}_{ij} &=& G \delta^{ij} \left( \vec{\theta}^2 - 
  (\vec{n} \cdot \vec{\theta})^2 \right) h + G \left(
  \theta^i n^j + \theta^j n^i \right) (\vec{n} \cdot \vec{\theta})
\,\, h + \alpha n^i n^j \vec{\theta}^2 h. 
\eea
The $\alpha$ dependent terms are proportional to the homogeneous
solution (\ref{5.8}) and can be absorbed in it so we set $\alpha$ to
zero. Then the noncommutative contribution to the metric at the
linearized level is 
\bea
\label{5.14}
g_{00} &=& 1 + \left( 1 + 3 G \vec{\theta}^2  - 
G (\vec{n} \cdot \vec{\theta})^2 \right) h, \nn
g_{0i} &=& 0, \nn
g_{ij} &=& - \delta^{ij} + \left[ n^i n^j + G \delta^{ij}
  \left( \vec{\theta}^2 - (\vec{n} \cdot \vec{\theta})^2 \right) +
  G \left(\theta^i n^j + \theta^j n^i \right) \vec{n} \cdot
  \vec{\theta} \right] h.
\eea

The effect of noncommutativity on a test particle can be seen by
analyzing the geodesic equation. As usual, the time component equation
relates the particle proper-time $\tau$ with the time $x^0 = t$ 
\be
\label{5.15}
\frac{dt}{d\tau} = 1 - h_{00} - \overline{h}_{00}.
\ee
Then disregarding terms quadratic in the velocity we read off the 
the potential from the space components of the geodesic equation 
\be
\label{5.16}
\frac{d^2 x^i}{dt^2} = - \frac{1}{2} \partial_i \left[ h + 
G \left( 3 \vec{\theta}^2 - (\vec{n} \cdot \vec{\theta} )^2 \right) h
\right]. 
\ee
The Newtonian potential has a noncommutative contribution
proportional to $\vec{\theta}^2$ which can be 
regarded as giving rise to an effective Newton
constant $G(1 + \frac{3}{2} G \vec{\theta}^2)$. It can be larger or smaller
than $G$ (recall that in the action (\ref{4.1}) we can set $\theta^2 
\rightarrow - \theta^2$). The angular dependent piece $(\vec{n} \cdot
\vec{\theta} )^2$ also contributes to the potential. The force on a test
particle is given by 
\be
\label{5.17}
\frac{d^2 x^i}{dt^2} = \left[ n^i + 3 G \left(
    \vec{\theta}^2 - (\vec{n} \cdot \vec{\theta})^2 \right) n^i + 
2 G (\vec{n} \cdot \vec{\theta}) \theta^i \right] \frac{h}{2r}.
\ee
Even though the term proportional to $\vec{\theta}^2 - (\vec{n} \cdot
\vec{\theta})^2$ looks like a dipole the force still goes like
$1/r^2$ and is radial. This radial correction has a strength given by
the effective Newton constant discussed before. And the term  
proportional to $\vec{\theta}$ produces in general a force off the plane of the
orbit and is also periodic for closed orbits. Notice that if $\vec{\theta}$ is
perpendicular to the plane of the orbit then no periodic effect due to
noncommutativity is seen. A better treatment, which could be confronted
which experimental data, would require to keep general relativity
corrections which  were disregarded in this approximation. 

\section{Conclusion}

We have shown how we can generalize properties of the
noncommutative parameter  $\theta^{\mu\nu}$ from flat spacetime to
curved spacetime in the simplest way possible. It can regarded as a
covariantly constant tensor or as satisfying (\ref{2.8}). In the case
of a covariantly constant $\theta^{\mu\nu}$ general diffeomorphisms reduce to
volume preserving transformations which also preserve
$\theta^{\mu\nu}$. This gives a natural explanation for this symmetry
which was found in \cite{Calmet:2005qm}. 

Simple extensions of the Moyal product or the Kontsevich product to
curved spacetime lead to nonassociative products. Maybe this is related to
the findings of \cite{Alvarez-Gaume:2006bn} where a structure which
can not be written in terms of the Moyal product was detected. This
means that more general deformations, may be along the lines of
\cite{Fedosov}, must be considered.

Another interesting aspect is that the propagator in curved spacetime
now depends on $\theta$ in contradistinction with the flat spacetime
case where propagators receive no noncommutative contributions. 

In the case of a covariantly constant $\theta^{\mu\nu}$ general
relativity reduces to unimodular gravity. In the linearized limit we
found a constraint between the time-space and space-space components of
$\theta^{\mu\nu}$. We then obtained the Newtonian
potential in unimodular gravity and its noncommutative correction. It
still depends on $1/r$ but has an effective Newton constant and an 
angular dependent piece. We have worked in the linearized
approximation and general relativity contributions should also be
taken into account to be able to confront with
experiments. Cosmological corrections due to noncommutativity may also
be relevant. Studies in these directions are being developed. 

\section{Acknowledgments}

E.H. is supported by FAPESP through grant 03/09044-9. V.O.R. is partially
supported by CNPq, FAPESP and PRONEX under contract CNPq 66.2002/1998-99. 
V.O.R. also thanks conversations with \'Alvaro Restuccia which pointed 
out reference \cite{Fedosov}.

\end{document}